\documentclass[10pt,conference]{IEEEtran}
\IEEEoverridecommandlockouts

\usepackage{cite}
\usepackage{hyperref}
\usepackage{mathtools}
\usepackage{amsmath,amssymb,amsfonts}
\usepackage{graphicx}
\usepackage{textcomp}
\usepackage{soul}
\usepackage[table]{xcolor}
\usepackage{etc}
\usepackage{algorithm}
\usepackage{algpseudocode}
\usepackage{booktabs}
\usepackage[normalem]{ulem}
\usepackage{subcaption}
\usepackage{enumitem}
\usepackage{balance}

\usepackage{multirow,array}

\newcolumntype{C}[1]{>{\centering\arraybackslash}p{#1}}

\definecolor{darkgreen}{rgb}{0.0, 0.5, 0.0}

\newcommand{\tool}{\textsc{Dr.}\,\textsc{Greens}\xspace}

\def\BibTeX{{\rm B\kern-.05em{\sc i\kern-.025em b}\kern-.08em
    T\kern-.1667em\lower.7ex\hbox{E}\kern-.125emX}}

\definecolor{codegreen}{rgb}{0.1,0.7,0.1}
\definecolor{codegray}{rgb}{0.5,0.5,0.5}

\newtcolorbox{rqbox}[1][]{
    colback=green!10,
    colframe=codegreen,
    arc=1mm,
    boxrule=0.5pt,
    coltitle=black,
    fonttitle=\bfseries,
    title=#1
}

\newtcolorbox{rqboxgray}[1][]{
    colback=codegray!10,
    colframe=codegray,
    arc=1mm,
    boxrule=0.5pt,
    coltitle=black,
    fonttitle=\bfseries,
    title=#1
}

\definecolor{green}{rgb}{0.0, 0.5, 0.0}

\begin{document}

\title{Identifying and Repairing Energy Smells with Agent Skills
}

\author{
\IEEEauthorblockN{Mohammadjavad Mehditabar, Saurabhsingh Rajput, Tushar Sharma}
\IEEEauthorblockA{\textit{Dalhousie University, Halifax, Canada} \\
\{javad,saurabh,tushar\}@dal.ca}}

\maketitle

\begin{abstract}

Energy smells are inefficient implementations that waste energy. Once code reaches production, these smells become costlier and riskier to fix, so energy efficiency checks should \emph{shift left}, \ie{} move earlier in the development lifecycle, to where code is written. Existing energy smell tools rely on fixed static rules and do not measure energy savings, whereas LLM-based optimizers mostly target execution time and rarely ground their changes in concrete smell definitions. To address both limitations, we present \tool{}, an agent skill that detects and repairs energy smells in Python code within the developer's coding environment. The skill grounds the coding agent in our taxonomy of 12 energy smells and 65 root causes, augmented with repair guidance drawn from 230 measured cases.
The skill also encapsulates a static analyzer and an independent verifier to confirm behavior preservation and energy saving for each attempted repair. We evaluated \tool{} in 18 sessions on three open-source repositories, with two models and three skill configurations. Of the 258 measured repairs, 68.6\% saved energy, with a median saving of 24.9\%. Compared with a baseline skill that lacks the taxonomy, the full skill found more energy smells with verified savings in all three repositories with both models. \tool{} helps developers detect energy hotspots and fix them reliably without altering program behavior, which makes energy-efficient development more feasible.

\end{abstract}

\section{Introduction}
\label{sec:intro}
Software runs on computational resources that consume electricity and the consumption is steadily increasing.
Data centers' electricity demand grew by 17\% in 2025 and is projected to double by 2030~\cite{iea2026keyquestions}. 
Inefficient code patterns waste energy that may seem to cost little in a single run, but scale up to a significant waste across millions of executions~\cite {mehditabar2026validated}. 
Such wasteful code patterns are \textit{energy smells}~\cite{vetro2013definition,mehditabar2026validated}. 

However, unlike traditional code smells
energy smells cannot be detected and resolved reliably statically alone.
Whether a pattern wastes energy depends on several factors including hardware, 
input workload~\cite{mehditabar2026validated,rajput2026energy}, the measurement methods~\cite{rajput2024enhancing,rajput2026codegreen}.
Detecting and repairing energy smells therefore requires measurements under realistic inputs.

Despite the importance of energy-efficient software, only 18\% of 122 surveyed programmers reported considering energy-efficiency during implementation~\cite{pang2015programmers}.
As a result, inefficiencies accumulate unchecked~\cite{maia2020debitum}, and AI-generated code makes this even worse~\cite{rajput2026beyond}. Once code reaches production, similar to bugs in software systems, energy smells are costlier and riskier to fix, however, often de-prioritized over more prominent bugs and issues~\cite{maia2020debitum,pang2015programmers,li2023combatting}.
In this paper, we emphasize that energy-efficiency checks should therefore \emph{shift left}, \ie{} move earlier in the development lifecycle, so that inefficiencies are detected and corrected earlier before they become part of the production. Given developers' increasing use of LLM-based coding agents~\cite{li2026aidev}, implementing energy-efficiency checks for coding agents is desirable and rational.

Existing approaches leave four key gaps. First, static analysis tools rely on pre-defined, hand-written rules~\cite{cruz2017leafactor,ribeiro2021ecoandroid,le2022ecocode,creedengo}.
Due to that, they cannot reason about runtime context or verify energy savings. 
Agentic refactoring approaches that rely on static analyzers inherits these weaknesses and can even degrade energy efficiency~\cite{dirocco2026automated}. Second, LLM-based optimizers mainly target execution time~\cite{zhao2026semopt,bai2025polo,deng2026perfagent,wang2026effiskill,lin2026eco}, which does not consistently predict energy~\cite{mehditabar2026validated}. 
Third, most approaches, such as EffiSkill~\cite{wang2026effiskill} and ECO~\cite{lin2026eco}, lack an iterative optimization mechanism guided by measured energy consumption. 
LASSI-EE~\cite{dearing2025leveraging} measures energy, but through a fixed workflow for GPU-related code.
Finally, most approaches target code optimization at a small scope. EffiSkill and EnergAgent~\cite{scheffler2025energagent} target competitive programming problems, refactor single scripts or functions.
Therefore, these approaches cannot directly spot and refactor inefficiencies in real-world code.

We present \tool{}, an agent skill~\cite{anthropic2025agentskills} for Claude.
A \textit{skill} is a reusable unit of documented expertise that gives a coding agent domain knowledge, a working procedure, and executable tools~\cite{anthropic2025agentskills}.
It runs inside the developer's coding agent and activates in response to a developer prompt, even when the developer does not explicitly ask for it. Our skill grounds the agent in our published taxonomy of 12 energy smells and 65 root causes~\cite{mehditabar2026validated}.
The agent flags suspected smells, confirms each against the repair reference for its root cause, and repairs the confirmed ones, running and measuring code at every step. 
It submits each repair to a verifier that checks behavior against the original and measures both versions, on inputs from a separate generator~\cite{shetty2026gso,peng2025coffe} so that the agent cannot tailor them to its repairs. After the session, we re-measure every repair and recommend only those that preserve behavior and save energy. On three open-source Python repositories with two models, 68.6\% of the measured repairs save energy, with a median saving of 24.9\%, a promising result given how context and hardware-dependent energy is. Our results also show that adding the taxonomy and the static analyzer to the skill leads the agent to find more energy smells.

In summary, this paper main contribution is \textbf{\tool{}}, to the best of our knowledge the first agent skill that detects and repairs energy smells in Python code, grounded in our published energy smell taxonomy. It equips the agent with our detailed \textbf{knowledge cards}, a detection and repair guide for each of the 65 root causes, built on 230 measured before-and-after cases, and with a \textbf{static analyzer} of 40 new rules and 31 Ruff rules that covers 49 root causes. We make the skill, the analyzer, and the measurement data accessible publicly in a replication package~\cite{drgreens2026repo}.

\section{\tool{}}
\label{sec:methodology}

\subsection{Overview}
\label{subsec:overview}

\begin{figure}[ht]
      \centering
      \includegraphics[width=\columnwidth]{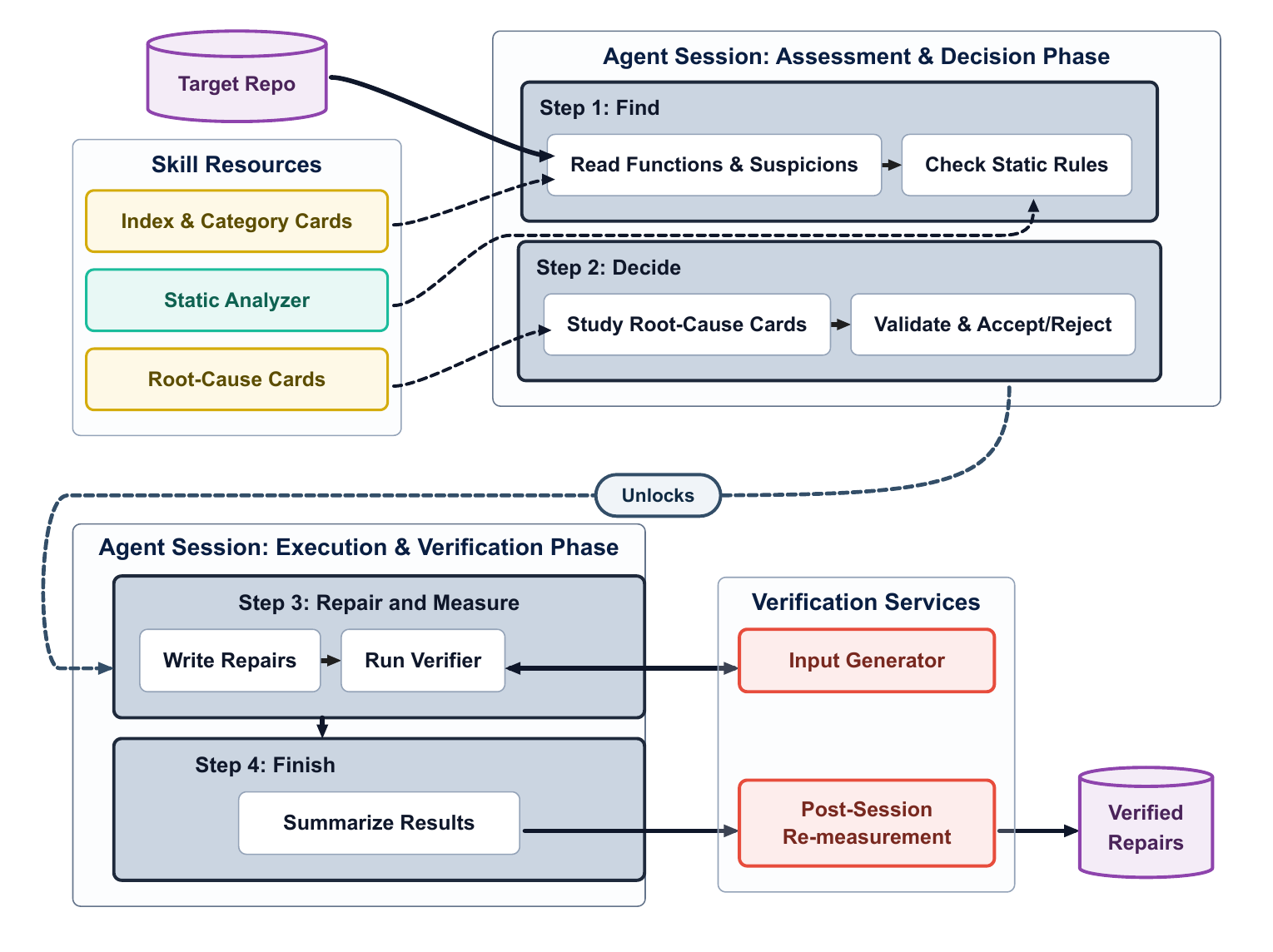}
      \caption{Overview of a \tool{} session.}
      \vspace{-4mm}
      \label{fig:pipeline_architecture}
\end{figure}

\tool{} finds energy smells in a given repository and repairs them so that the repaired code uses less energy and preserves the original behavior. We target repositories mainly written in Python because it is among the most used programming languages~\cite{github2024octoverse}, yet among the least energy-efficient~\cite{pereira2021ranking}.
The skill grounds the agent in our taxonomy~\cite{mehditabar2026validated}, which defines $12$ energy smells, \ie{} categories of inefficient code, and $65$ root causes, \ie{} the specific code choices that cause the waste. 
\tool{} treats functions as the unit of analysis, and each repair replaces one function and addresses one root cause.
Figure~\ref{fig:pipeline_architecture} shows a session and the steps the agent follows. We measure energy with CodeGreen~\cite{rajput2026codegreen}, a profiler for fine-grained, low-noise energy measurement.

\subsection{Design Rationale}
\label{subsec:rationale}

Four observations shape the design. \textit{First}, the energy savings of a change depends on the hardware, the runtime, and the workload, so only running the code can establish it. We therefore build \tool{} for a coding agent, which can execute code during its analysis, and give it an \textit{energy probe} that measures any function on inputs of its choice.
\textit{Second}, models without guidance struggle to recognize energy-inefficient code~\cite{rajput2026beyond,apsan2026generating,gulu2026balancing}, while fixed workflows restrict what an agent can explore~\cite{dearing2025leveraging}. An agent skill~\cite{anthropic2025agentskills} balances the two: it prescribes the steps and supplies the taxonomy, and the agent decides how to read, run, and test the code within each step. The agent loads reference files only when a step needs them, so the $65$ detailed cards do not fill its context unnecessarily.
\textit{Third}, an agent that reads thousands of functions can overlook a pattern, while a static rule checks every function but cannot judge context. We therefore add a static analyzer whose findings the agent reviews.
\textit{Fourth}, an agent that measures its own repairs could choose inputs that favor them. We therefore keep the evidence outside the agent's control. The agent never edits the repository; a separate \textit{input generator} creates its measurement inputs, a verifier checks and measures every repair, and we measure every repair again after the session. The developer can then apply the recommended repairs to their codebase.

\subsection{Reference Knowledge and Static Analyzer}
\label{subsec:knowledge}

Following the progressive disclosure design of agent skills~\cite{anthropic2025agentskills}, the reference knowledge has three levels of increasing detail. A \textit{taxonomy index} gives a one-line description of what to look for under each root cause. Twelve \textit{category cards}, one per energy smell, give a one-line repair for each of its root causes. Sixty-five \textit{root cause cards} give the detail needed to confirm or reject a suspicion. Each holds the definition and example of its root cause, taken verbatim from the taxonomy, a description of how the waste appears in code, measured repair cases, and near-misses, \ie{} code that resembles the root cause but where the repair would be wrong.

The taxonomy comes from studies across languages and whether a repair pays off depends on context. Two authors with more than five years of software engineering experience wrote $230$ repair cases that cover all $65$ root causes, with two to seven cases per root cause. Each case consists of a function with the waste (\textit{before}), an equivalent function with the repair (\textit{after}), and inputs.
We measure each case with a \textit{paired protocol}. Both versions first run on every input, and only versions with identical outputs proceed to measurement. Each version then runs the same number of calls, about three seconds of work, so that their energy readings are large enough to measure and directly comparable. We report the mean over nine pairs per input. An input counts as saving or costing energy only if the 95\% confidence interval of the paired difference (Student's $t$ over the nine pairs) excludes zero and the difference exceeds the \textit{noise floor}, which we obtain by measuring a version against an identical copy of itself.
Of the $230$ cases, $146$ save energy on some input and cost more on none. For $59$ of the $65$ root causes, at least one case saves energy without costing more on any input. A case that costs more does not refute its root cause, because our cases cover only a few forms of it. The cards therefore report each case with its per-input result and leave the judgment to the agent.
Our static analyzer combines $40$ rules over Python's syntax tree with $31$ Ruff rules~\cite{astral_ruff_rules} whose meaning implies a specific root cause. Each rule maps to one or more root causes, and together the rules cover $49$ of the $65$. Whether a pattern wastes energy depends on context, so the analyzer reports observed patterns without judging them, \eg{} ``a string is grown with \texttt{+=} inside a loop at line 45'', each tagged with the related root causes. To validate the rules, two authors wrote $515$ test cases, $319$ of which the analyzer must not report, \ie{} false positive cases.

\subsection{Workflow}
\label{subsec:workflow}

\subsubsection{Step 1 (Find)}
\label{subsec:step1}

Step~1 favors recall. The agent reads every function and records each suspicion with the root cause it resembles, the lines involved, and its reason. On a large repository, the agent may hand over groups of files to subagents, which keeps each context small. In this step, the agent sees only the taxonomy index and the category cards. When the analyzer is present, Step~1 ends by comparing its findings with the agent's list. For each function or root cause that the analyzer flags and the agent missed, the agent reads the code and decides whether to add it.

\subsubsection{Step 2 (Decide)}
\label{subsec:step2}

Step~2 favors precision. The root cause cards become available at this step, and the agent examines every listed function in depth. It opens the card of each suspicion. It accepts the suspicion only if the card's definition matches the code and the repair would preserve behavior on every input it considers, including edge cases. The near-misses on the cards help it reject code that resembles a root cause but where the repair would be wrong.

\subsubsection{Step 3 (Repair and Measure)}
\label{subsec:step3}

Finishing Step~2 gives the agent access to the input generator and the verifier. Since each measurement takes time, the skill limits how many repairs a session can measure. The agent repairs the accepted suspicions it expects to save the most and records the rest as \textit{deferred}.
\textbf{Inputs.} The agent requests inputs for the functions it plans to repair, and the input generator creates them with the \textsc{DeepSeek V4.1} Flash model. For each function, it produces a few small \textit{edge inputs} to check behavior, and four high-load \textit{measured inputs}, which make the function run long enough for its energy to exceed the noise. Following GSO~\cite{shetty2026gso}, we keep an input only if it runs on the original code and gives the same output twice, because a nondeterministic output cannot confirm equivalence. Following COFFE~\cite{peng2025coffe}, the LLM writes eight candidate measured inputs, and we keep the four that use the most energy on the original code and together cover at least $40\%$ of the function's statements.
\textbf{Verification.} The agent submits each repair as the full new source of the function, with its root cause. The verifier runs every edge and measured input through both versions and compares the return value, the state of the arguments, the printed output, and the files written. Any difference makes the repair \textit{not equivalent}. Otherwise, the verifier measures both versions on the measured inputs with the paired protocol (Section~\ref{subsec:knowledge}).
\textbf{Verdicts.} A repair is \textit{confirmed} when it saves energy on all four inputs, and \textit{partial} when it saves on some and changes nothing on the rest. It is \textit{mixed} when it saves on some and costs more on others, \textit{regressed} when every input that changes costs more, and \textit{negligible} when no input changes beyond a noise floor of at most $3\%$. When the noise floor exceeds $3\%$, a repair is \textit{inconclusive}, unless all four inputs save energy or at least one costs more. Only the five measured verdicts count toward the session's limit, and a measured verdict is final. If a repair is not equivalent or fails to run, the agent may fix and resubmit it.
\textbf{Re-measurement.} After the session, we measure every attempted repair again with the verifier, so that the reported results do not depend on the agent's own measurements. Only one repair per function can be applied, so among the confirmed repairs of a function we recommend the one with the highest mean saving. The agent then ends the session with a summary that lists each repair with its root cause and verdict (Step~4, \textit{Finish}).

\section{Evaluation}
\label{sec:evaluation}

\textbf{Setup.} We compare three skills of \tool{} to understand the contribution of each knowledge layer. The \unguided skill includes the energy probe, the input generator, and the verifier, but omits the taxonomy, so the agent judges waste from its own knowledge. The \taxo skill adds the taxonomy index, the category cards, and the root cause cards, and the \taxorules skill further adds the static analyzer. We run the skills in Claude Code with permission to run any command. We use two models of different capability and cost, Claude Sonnet 5.5 and Claude Haiku 5.5, both at high reasoning effort.
We sample Python repositories with SEART GitHub Search~\cite{Dabic:msr2021data}, requiring at least $100$ stars, $500$ commits, and a push after Sept~12,~2026. We screen the first ten repositories and retain those for which package installation succeed and the code runs successfully. Three repositories meet with all criteria: \texttt{vespa-engine/sample-apps} ($443$ functions), \texttt{dnanexus/dx-toolkit} ($2{,}543$ functions), and \texttt{open-telemetry/community} ($95$ functions). We run one session per repository, skill, and model, \ie{} $18$ sessions. Each session runs independently 
on a dedicated server with two Intel Xeon E5-2650 processors. 
The verifier serves as the oracle: a repair is correct when it preserves behavior and saves energy.

\textbf{Results.} Table~\ref{tab:rq1-runs} reports each session. A \textit{smelly function} is a function with at least one accepted smell. A repair \textit{saves energy} when its verdict is confirmed or partial.

\begin{table}[t]
\centering
\caption{Results per session. C = \unguided, T = \taxo, F = \taxorules. Bold marks the best skill per repo. and model.}
\label{tab:rq1-runs}
\footnotesize
\setlength{\tabcolsep}{2pt}
\begin{tabular}{@{}llrrrrrr@{}}
\toprule
 & & \multicolumn{3}{c}{\textbf{Sonnet}} & \multicolumn{3}{c}{\textbf{Haiku}} \\
\cmidrule(lr){3-5}\cmidrule(l){6-8}
 & & Smelly & Saving/ & Median & Smelly & Saving/ & Median \\
\textbf{Repo} & \textbf{Skill} & (\%) & meas. & (\%) & (\%) & meas. & (\%) \\
\midrule
\multirow{3}{*}{\texttt{sample-apps}}
 & C   & 39 (8.8)            & 12/20          & \textbf{33.2} & 41 (9.3)           & 11/15          & 23.3 \\
 & T   & 83 (18.7)           & \textbf{16}/20 & 19.8          & \textbf{81} (18.3) & \textbf{16}/20 & \textbf{26.3} \\
 & F & \textbf{102} (23.0) & 13/20          & 15.1          & 56 (12.6)          & 15/20          & 24.5 \\
\midrule
\multirow{3}{*}{\texttt{dx-toolkit}}
 & C   & 46 (1.8)            & 13/20          & \textbf{56.7} & 57 (2.2)           & 1/5            & \textbf{99.7}\\
 & T   & 83 (3.3)            & \textbf{14}/20 & 40.8          & 70 (2.8)           & 6/9            & 48.3 \\
 & F & \textbf{178} (7.0)  & 10/20          & 29.6          & \textbf{94} (3.7)  & \textbf{10}/12 & 39.5 \\
\midrule
\multirow{3}{*}{\texttt{community}}
 & C   & 9 (9.5)             & 7/8            & \textbf{23.1} & 11 (11.6)          & \textbf{4}/6   & \textbf{21.6} \\
 & T   & 16 (16.8)           & 10/14          & 10.8          & \textbf{15} (15.8) & \textbf{4}/7   & 10.3 \\
 & F & \textbf{21} (22.1)  & \textbf{12}/17 & 5.1           & 13 (13.7)          & 3/5            & 18.3 \\
\bottomrule
\end{tabular}
\vspace{-4mm}
\end{table}

\paragraph{Most repairs save energy}
The agents submitted $317$ repairs, of which $258$ received a measured verdict. Our re-measurement shows that $68.6\%$ of these save energy (95\% CI $63.3\%$ to $73.2\%$), with a median saving of $24.9\%$ (95\% CI $17.8\%$ to $26.7\%$). Both intervals come from a percentile bootstrap that resamples the $18$ sessions $10{,}000$ times. Sonnet and Haiku save energy at similar rates, $67.3\%$ and $70.7\%$ of their measured repairs, so the share of verified repairs holds across models of different capability.

\paragraph{The taxonomy widens what the agent finds}
Both taxonomy skills find more smelly functions than \unguided on every repository and with both models. The static analyzer raises the count in four of the six combinations, including all three with Sonnet, adding $1$ to $17$ functions per session. With either taxonomy skill, Sonnet finds more smelly functions than Haiku on every repository, while with \unguided Haiku finds more. To test whether this gap reflects model capability, we also ran the three skills with Claude Opus 5.5 on sample-apps. 
Opus finds the most smelly functions with every skill, and the taxonomy roughly doubles its count ($146$ with \taxo and $131$ with \taxorules, against $69$ with \unguided), while producing at least as many saving repairs as Sonnet and Haiku.

\noindent
\paragraph{Wider search adds smaller savings}
\unguided has the highest median savings in five of the six combinations of repository and model, but it produces fewer saving repairs ($48$, against $66$ for \taxo and $63$ for \taxorules).
On functions that two skills repair with the same model, their best savings are about the same (median difference $0$). With Haiku, the three medians are close ($25.1\%$ to $26.6\%$). With Sonnet, the taxonomy skills also repair $36$ functions that \unguided misses, mostly with small savings (median $8.2\%$).
The static analyzer shows the same pattern. Because several sessions often write the same edit, we count each distinct edit once, which leaves $192$ edits, $65.6\%$ of which save energy. However, the analyzer flags only $17.9\%$ of all functions; $100$ of these edits change a flagged function. Edits to flagged functions save energy more often than other edits ($69.0\%$ vs $62.0\%$), but their median saving is $17.5$ percentage points lower (95\% CI $2.6$ to $38.4$, bootstrap over edits). 
The knowledge layers therefore let the agent catch smaller savings that it would otherwise miss.

\begin{rqbox}
\textbf{Takeaway.} Of $258$ measured repairs, \tool{} optimizes $68.6\%$ cases, with a median saving of $24.9\%$.
\end{rqbox}

\section{Implications}
\label{sec:implications}

\tool{} has implications for practice and research. \textit{Developers} can use it within their agentic workflows to keep their code energy-efficient. Reviewing measured repairs replaces searching for waste manually, and the repairs show which coding habits waste energy. \textit{Organizations} can treat energy as a quality attribute during development, not after release. The static analyzer can run on every commit, and the full skill can run before each release. \textit{Tool builders} can extend the approach to other languages or adapt it to other quality attributes, such as performance or memory use. \textit{Model developers} can use the verified repairs as training data for code models that generate efficient code from the start, and as a benchmark for energy-aware agents. \textit{Researchers} can use the skill to study what models know about energy without guidance and how agents decide what to repair. Runs on more repositories can show which energy smells are most common in which domains.

\section{Related Work}
\label{sec:related}

\textbf{Energy smells.}\label{sec:rw-energy-smells}
Early work measured the effect of energy smells on an embedded system~\cite{vetro2013definition} and proposed removing them with reengineering services~\cite{gottschalk2012removing}. Later catalogs targeted mobile apps~\cite{cruz2019catalog} and the web~\cite{rani2024energy}, each covering one domain or platform. Our prior work~\cite{mehditabar2026validated} addresses this with a language-agnostic taxonomy of 65 root causes under 12 energy smells.

\textbf{Performance inefficiencies.}\label{sec:rw-performance}
Earlier work catalogs performance antipatterns~\cite{smith2000software} and finds them with rule-based checkers~\cite{jin2012understanding} and linters such as Perflint and Ruff~\cite{shaw_perflint,astral_ruff_rules}, and learned models~\cite{zhao2024enhancing,garg2022deepdev}. LLM-based optimizers guide generation with mined antipatterns or optimization rules~\cite{lin2026eco,zhao2026semopt}, package slow-to-fast transformations as agent skills~\cite{wang2026effiskill}, or give a coding agent profiler feedback and a verifier across a repository, as PerfAgent does~\cite{deng2026perfagent}. All of them target performance, although faster is not always more energy-efficient~\cite{mehditabar2026validated}.

\textbf{Energy inefficiency tools.}\label{sec:rw-energy-detection}
Most energy smell detectors target Android, either by running the app~\cite{liu2014greendroid} or by static analysis~\cite{hecht2015detecting}, and some also refactor the smells they find~\cite{cruz2017leafactor,ribeiro2021ecoandroid}. ecoCode~\cite{le2022ecocode}, now creedengo~\cite{creedengo}, brings such rules to SonarQube for several languages. Among LLM-based approaches, Di Fiore \etal{}~\cite{dirocco2026automated} let a team of LLM agents fix the smells that SonarQube plugins report in Java classes. LASSI-EE~\cite{dearing2025leveraging} and EnergAgent~\cite{scheffler2025energagent} use energy measurements feedback to refine parallel and LLM-generated C++ code.

\textbf{Energy efficiency of LLM-generated code.}\label{sec:rw-llm}
Studies comparing code generated by different LLMs and configurations~\cite{solovyeva2025ai,ashraf2025energy} find that asking for energy-efficient code does not reliably save energy~\cite{apsan2026generating}. Rajput \etal{}~\cite{rajput2026beyond} instead train a code model on simulated energy feedback.
\section{Threats to Validity}
For \textbf{construct validity}, the four inputs on which a repair is measured may not represent the data a unit receives in practice. We mitigate this by generating most high-load test cases that impact energy, keeping only inputs that cover at least 40\% of its statements.
For \textbf{internal validity}, 
energy readings may be distorted by noise. We mitigate this by running one session at a time on a dedicated machine, alternating both versions in one process inside their own repository, and counting only differences above a measured noise floor. 
\textbf{External validity} threats the generalizability of the results. 
We mitigate this by sampling repositories across domains and sizes and keeping the skill to plain instructions and scripts any model can follow.

    \section{Conclusions}
\label{sec:conclusion}

\tool{} detects energy smells and repairs them with measured savings, by packaging an energy smell taxonomy, a static analyzer, and an independent verifier as an agent skill that runs inside the developer's coding environment. In an evaluation across $18$ distinct combinations of repositories, models, and skills, \tool{} generated $258$ measureable repairs; $68.6$\% successfully yielded energy savings, with a median energy saving of $24.9$\%.
By bringing energy checks to where code is written, \tool{} gives developers practical support to measure how green their software is and to fix energy smells through automated refactoring across domains.

\bibliographystyle{IEEEtran} 
\balance
\bibliography{references}

@article{pang2015programmers,
  title={What do programmers know about software energy consumption?},
  author={Pang, Candy and Hindle, Abram and Adams, Bram and Hassan, Ahmed E},
  journal={IEEE Software},
  year={2015},
}

@inproceedings{maia2020debitum,
  title={E-debitum: managing software energy debt},
  author={Maia, Daniel and Couto, Marco and Saraiva, Jo{\~a}o and Pereira, Rui},
  booktitle={Proceedings of the 35th IEEE/ACM International Conference on Automated Software Engineering},
  year={2020}
}

@inproceedings{apsan2026generating,
  title={Generating Energy-Efficient Code via Large-Language Models-Where are we now?},
  author={Apsan, Radu and Stoico, Vincenzo and Albonico, Michel and Dhar, Rudra and Vaidhyanathan, Karthik and Malavolta, Ivano},
  booktitle={Proceedings of the 2026 IEEE/ACM 48th International Conference on Software Engineering},
  year={2026}
}

@inproceedings{ashraf2025energy,
  title={Energy-aware code generation with llms: Benchmarking small vs. large language models for sustainable ai programming},
  author={Ashraf, Humza and Danish, Syed Muhammad and Leivadeas, Aris and Otoum, Yazan and Sattar, Zeeshan},
  booktitle={2025 3rd International Conference on Foundation and Large Language Models (FLLM)},
  year={2025},
  organization={IEEE}
}

@inproceedings{gulu2026balancing,
  title={Balancing Green and Clean Code: Prompting for LLM-Based Refactoring},
  author={Gulu-zada, Feray and Anwar, Hina},
  booktitle={20th International Symposium on Empirical Software Engineering and Measurement (ESEM 2026)},
  year={2026},
  organization={Schloss Dagstuhl--Leibniz-Zentrum f{\"u}r Informatik}
}

@article{rajput2026beyond,
  title={Beyond the Need for Speed: Energy-Aware Code Generation via Simulation-Guided Reinforcement Learning},
  author={Rajput, Saurabhsingh and Sharma, Tushar},
  journal={arXiv preprint arXiv:2607.04577},
  year={2026}
}

@article{deng2026perfagent,
  title={PerfAgent: Profiler-Guided Iterative Refinement for Repository-Level Code Optimization},
  author={Deng, Ryan and Liu, Yuanzhe and Lipka, Bastian and Ma, Yao and Chen, Xuhao and Kaler, Tim and Ganhotra, Jatin},
  journal={arXiv preprint arXiv:2607.19653},
  year={2026}
}

@article{wang2026effiskill,
  title={Effiskill: Agent skill based automated code efficiency optimization},
  author={Wang, Zimu and Shi, Yuling and Li, Mengfan and Liu, Zijun and Zhang, Jie M and Wan, Chengcheng and Gu, Xiaodong},
  journal={arXiv preprint arXiv:2603.27850},
  year={2026}
}

@inproceedings{bai2025polo,
  title={POLO: An LLM-Powered Project-Level Code Performance Optimization Framework.},
  author={Bai, Jiameng and Xu, Ruoyi and Wu, Sai and Yang, Dingyu and Zhao, Junbo and Chen, Gang},
  booktitle={IJCAI},
  year={2025}
}

@article{zhao2026semopt,
  title={SemOpt: LLM-Driven Code Optimization via Rule-Based Analysis},
  author={Zhao, Yuwei and Xiao, Yuan-An and Xiao, Qianyu and Zhang, Zhao and Xiong, Yingfei},
  journal={ACM Transactions on Software Engineering and Methodology},
  year={2026},
}

@inproceedings{lin2026eco,
  author    = {Lin, Hannah and Maas, Martin and Roquemore, Maximilian and Hasanzadeh, Arman and Lewis, Fred and Simonson, Yusuf and others},
  title     = {{ECO}: An {AI}-Driven Code Efficiency Optimizer for Warehouse Scale Computers},
  booktitle = {20th USENIX Symposium on Operating Systems Design and Implementation (OSDI)},
  year      = {2026}
}

@inproceedings{cruz2017leafactor,
  title={Leafactor: Improving energy efficiency of android apps via automatic refactoring},
  author={Cruz, Luis and Abreu, Rui and Rouvignac, Jean-No{\"e}l},
  booktitle={2017 IEEE/ACM 4th International Conference on Mobile Software Engineering and Systems (MOBILESoft)},
  year={2017},
  organization={IEEE}
}

@inproceedings{le2022ecocode,
  title={Ecocode: A sonarqube plugin to remove energy smells from android projects},
  author={Le Goaer, Olivier and Hertout, Julien},
  booktitle={Proceedings of the 37th IEEE/ACM International conference on automated software engineering},
  year={2022}
}

@inproceedings{mehditabar2026validated,
  title={Watts This Smell: A Comprehensive Taxonomy of Software Energy Smells},
  author={Mehditabar, Mohammadjavad and Rajput, Saurabhsingh and Sharma, Tushar},
  booktitle={2026 IEEE International Conference on Software Maintenance and Evolution (ICSME)},
  year={2026},
  organization={IEEE}
}

@inproceedings{li2026aidev,
  title={Aidev: studying ai coding agents on github},
  author={Li, Hao and Zhang, Haoxiang and Hassan, Ahmed E},
  booktitle={Proceedings of the 23rd International Conference on Mining Software Repositories},
  year={2026}
}

@misc{anthropic2025agentskills,
  author       = {{Anthropic}},
  title        = {{Agent Skills Overview}},
  howpublished = {\url{https://platform.claude.com/docs/en/agents-and-tools/agent-skills/overview}},
  year         = {2025},
}

@inproceedings{gottschalk2012removing,
  title={Removing energy code smells with reengineering services},
  author={Gottschalk, Marion and Josefiok, Mirco and Jelschen, Jan and Winter, Andreas},
  booktitle={INFORMATIK 2012},
  year={2012},
  organization={Gesellschaft f{\"u}r Informatik eV}
}

@inproceedings{vetro2013definition,
  title={Definition, implementation and validation of energy code smells: an exploratory study on an embedded system},
  author={Vetro, Antonio and Ardito, Luca and Procaccianti, Giuseppe and Morisio, Maurizio and others},
  booktitle={Proceedings of ENERGY 2013: The Third International Conference on Smart Grids, Green Communications and IT Energy-aware Technologies},
  year={2013}
}

@techreport{iea2026keyquestions,
  author      = {{International Energy Agency}},
  title       = {Key Questions on Energy and {AI}},
  institution = {IEA},
  year        = {2026},
  month       = apr,
  url         = {https://www.iea.org/reports/key-questions-on-energy-and-ai},
  note        = {Licence: CC BY 4.0}
}

@article{rajput2024enhancing,
  title={Enhancing energy-awareness in deep learning through fine-grained energy measurement},
  author={Rajput, Saurabhsingh and Widmayer, Tim and Shang, Ziyuan and Kechagia, Maria and Sarro, Federica and Sharma, Tushar},
  journal={ACM Transactions on Software Engineering and Methodology},
  year={2024},
}

@inproceedings{rajput2026energy,
  title={Energy Flow Graph: Modeling Software Energy Consumption},
  author={Rajput, Saurabhsingh and Sharma, Tushar},
  booktitle={Proceedings of the 34th ACM International Conference on the Foundations of Software Engineering},
  year={2026}
}

@inproceedings{rajput2026codegreen,
  title={CodeGreen: Towards Improving Precision and Portability in Software Energy Measurement},
  author={Rajput, Saurabhsingh and Sharma, Tushar},
  booktitle={Proceedings of the 34th ACM International Conference on the Foundations of Software Engineering},
  year={2026}
}

@article{shetty2026gso,
  title={Gso: Challenging software optimization tasks for evaluating swe-agents},
  author={Shetty, Manish and Jain, Naman and Liu, Jinjian and Kethanaboyina, Vijay and Sen, Koushik and Stoica, Ion},
  journal={Advances in Neural Information Processing Systems},
  volume={38},
  year={2026}
}

@article{peng2025coffe,
  title={Coffe: A code efficiency benchmark for code generation},
  author={Peng, Yun and Wan, Jun and Li, Yichen and Ren, Xiaoxue},
  journal={Proceedings of the ACM on Software Engineering},
  number={FSE},
  year={2025},
}

@inproceedings{smith2000software,
  title={Software performance antipatterns},
  author={Smith, Connie U and Williams, Lloyd G},
  booktitle={Proceedings of the 2nd international workshop on Software and performance},
  year={2000}
}

@article{jin2012understanding,
  title={Understanding and detecting real-world performance bugs},
  author={Jin, Guoliang and Song, Linhai and Shi, Xiaoming and Scherpelz, Joel and Lu, Shan},
  journal={ACM SIGPLAN Notices},
  year={2012},
}

@misc{shaw_perflint,
  author       = {Anthony Shaw},
  title        = {Perflint: A {Python} Linter for Performance Anti-Patterns},
  howpublished = {\url{https://github.com/tonybaloney/perflint}},
  year         = {2022}
}

@misc{astral_ruff_rules,
  author       = {{Astral Software}},
  title        = {Ruff: Rules Reference},
  howpublished = {\url{https://docs.astral.sh/ruff/rules/}},
  year         = {2023}
}

@inproceedings{garg2022deepdev,
  title={Deepdev-perf: a deep learning-based approach for improving software performance},
  author={Garg, Spandan and Moghaddam, Roshanak Zilouchian and Clement, Colin B and Sundaresan, Neel and Wu, Chen},
  booktitle={ACM Joint European Software Engineering Conference and Symposium on the Foundations of Software Engineering},
  year={2022}
}

@inproceedings{zhao2024enhancing,
  title={Enhancing performance bug prediction using performance code metrics},
  author={Zhao, Guoliang and Georgiou, Stefanos and Hassan, Safwat and Zou, Ying and Truong, Derek and Corbin, Toby},
  booktitle={Proceedings of the 21st International Conference on Mining Software Repositories},
  year={2024}
}

@inproceedings{scheffler2025energagent,
  title={EnergAgent: Energy-Aware Code Refinement with LLM-Based Agents},
  author={Scheffler, Jonas and Westing, Max},
  booktitle={EnviroInfo 2025},
  year={2025},
  organization={Gesellschaft f{\"u}r Informatik eV}
}

@article{dearing2025leveraging,
  title={Leveraging llms to automate energy-aware refactoring of parallel scientific codes},
  author={Dearing, Matthew T and Tao, Yiheng and Wu, Xingfu and Lan, Zhiling and Taylor, Valerie},
  journal={arXiv preprint arXiv:2505.02184},
  year={2025}
}

@inproceedings{dirocco2026automated,
  author={Di Fiore, Guido Maria and Di Rocco, Juri and Muccini, Henry},
  title     = {Automated Energy Smell Refactoring of {Java} Code: An {LLM} Agents-Based Approach},
  booktitle = {Proceedings of the International Conference on the Quality of Information and Communications Technology (QUATIC)},
  year      = {2026}
}

@article{liu2014greendroid,
  author={Liu, Yepang and Xu, Chang and Cheung, S. C. and L{\"u}, Jian},
  journal={IEEE Transactions on Software Engineering}, 
   title={{GreenDroid}: Automated Diagnosis of Energy Inefficiency for Smartphone Applications},
  year={2014},
  doi={10.1109/TSE.2014.2323982}}

@phdthesis{hecht2015detecting,
  title={Detecting antipatterns in android apps},
  author={Hecht, Geoffrey and Rouvoy, Romain and Moha, Naouel and Duchien, Laurence},
  year={2015},
  school={INRIA Lille; INRIA}
}

@inproceedings{ribeiro2021ecoandroid,
  title={Ecoandroid: An android studio plugin for developing energy-efficient java mobile applications},
  author={Ribeiro, Ana and Ferreira, Jo{\~a}o F and Mendes, Alexandra},
  booktitle={2021 IEEE 21st international conference on software quality, reliability and security (QRS)},
  year={2021},
  organization={IEEE}
}

@inproceedings{rani2024energy,
  title={Energy patterns for web: An exploratory study},
  author={Rani, Pooja and Zellweger, Jonas and Kousadianos, Veronika and Cruz, Luis and Kehrer, Timo and Bacchelli, Alberto},
  booktitle={Proceedings of the 46th International Conference on Software Engineering: Software Engineering in Society},
  year={2024}
}

@article{cruz2019catalog,
  title={Catalog of energy patterns for mobile applications},
  author={Cruz, Luis and Abreu, Rui},
  journal={Empirical software engineering},
  year={2019},
}

@inproceedings{solovyeva2025ai,
  title={Ai-powered, but power-hungry? energy efficiency of llm-generated code},
  author={Solovyeva, Lola and Weidmann, Sophie and Castor, Fernando},
  booktitle={IEEE/ACM Second International Conference on AI Foundation Models and Software Engineering (Forge)},
  year={2025},
}

@misc{creedengo,
  author={{Green Code Initiative}},
  title={Creedengo: Rules Specifications for Energy-Efficient Code},
  howpublished={\url{https://github.com/green-code-initiative/creedengo-rules-specifications}},
  note={Accessed: November 2026}
}

@misc{github2024octoverse,
 author={{GitHub}},
 title={Octoverse: {AI} Leads {Python} to Top Language as the Number of Global Developers Surges},
 year={2024},
 howpublished={\url{https://github.blog/news-insights/octoverse/octoverse-2024/}},
 note={Accessed: November 2026}
}

@article{pereira2021ranking,
  title={Ranking programming languages by energy efficiency},
  author={Pereira, Rui and Couto, Marco and Ribeiro, Francisco and Rua, Rui and Cunha, J{\'a}come and Fernandes, Jo{\~a}o Paulo and Saraiva, Jo{\~a}o},
  journal={Science of Computer Programming},
  year={2021},
}

@inproceedings{Dabic:msr2021data,
  author    = {Ozren Dabic and Emad Aghajani and Gabriele Bavota},
  title     = {Sampling Projects in GitHub for {MSR} Studies},
  booktitle = {18th {IEEE/ACM} International Conference on Mining Software Repositories,
               {MSR} 2021},
  year      = {2021}
}

@article{li2023combatting,
  title={Combatting energy issues for mobile applications},
  author={Li, Xueliang and Chen, Junyang and Liu, Yepang and Wu, Kaishun and Gallagher, John P},
  journal={ACM Transactions on Software Engineering and Methodology},
  year={2023},
}

@software{drgreens2026repo,
author = {Mehditabar, Mohammadjavad and Rajput, Saurabhsingh and Sharma, Tushar},
title = {{Ctrl+Shift+Left: Detecting and Refactoring Energy Smells using Agent Skills}},
url = {https://github.com/SMART-Dal/Energy-Smell-Detection-with-Agents-Skills},
year = {2026}
}

\end{document}